\documentclass[12pt]{iopart}

\usepackage{iopams}

\usepackage{graphicx}
\usepackage{theorem}
\usepackage[dvipdfm]{hyperref}

\theoremstyle{plain}
\newtheorem{Thm}{Theorem}
\newtheorem{Lem}[Thm]{Lemma}
\newtheorem{Cor}[Thm]{Corollary}
\newtheorem{Prop}[Thm]{Proposition}

\theorembodyfont{\rmfamily}
\newtheorem{Proof}{Proof}

\begin{document}

\title{Vertex operator for the non-autonomous ultradiscrete KP equation}

\author{Yoichi Nakata}
\address{Graduate School of Mathematical Sciences, The University of Tokyo, 3-8-1 Komaba, Meguro-ku, 153-8914 Tokyo, Japan}
\ead{ynakata@ms.u-tokyo.ac.jp}
\begin{abstract}
We propose an ultradiscrete analogue of the vertex operator in the case of the ultradiscrete KP equation--several other ultradiscrete equations--which maps $N$-soliton solutions to $N+1$-soliton ones.
\end{abstract}

\vspace{2pc}

\bigskip

\section{Introduction}
The KP equation is widely considered as the paradigm of soliton equations. The main results of soliton theory, including first and foremost the celebrated Sato theory, were discovered in the study of this equation. The discovery of the vertex operator, which maps $N$-soliton solutions to $N+1$-soliton ones, figures prominently among these results.

The discrete KP equation (or Hirota-Miwa equation), which is a discretized version of the KP equation, is also regarded as a fundamental discrete soliton equation. By restricting its solutions it reduces to many well-known discrete soliton equations as, for example, the discrete KdV equation or the discrete Toda equation.

Soliton Cellular Automata form a class of cellular automata that exhibit soliton-like behaviour and possess a rich structure including the existence of explicit $N$-soliton solutions and an infinite amount of conserved quantities \cite{NTS}, like most ordinary soliton equations. The ``Box and Ball System" (BBS) \cite{TS} is the main representative of this class. It is related to the discrete soliton equations, through a limiting procedure called ``ultradiscretization" \cite{TTMS}.

The non-autonomous ultradiscrete KP equation is obtained by ultradiscretizing the non-autonomous discrete KP equation \cite{WTS}. Tokohiro \etal presented its $N$-soliton solution and described the dynamics of the BBS with several kinds of balls in \cite{TTM}. Shinzawa and Hirota discussed the consistency conditions of the B\"acklund transformation for the autonomous ultradiscrete KP equation in \cite{SH}.

Recently, these systems draw increasing interest due to the establishment of relationships to other mathematical topics, for example, to algebraic geometry and representation theory. It is therefore fruitful to clarify the symmetries and the algebraic structure of ultradiscrete soliton equations, as was done for the continuous ones. Takahashi and Hirota presented an approach based on so-called ``permanent type solutions" \cite{TH} (which are expressed as signature-free Casorati determinants) to discuss particular solutions of ultradiscrete systems. Nagai presented identities for permanent type solutions, which can be considered as ultradiscrete analogues of Pl\"ucker relations for determinants in \cite{Nh}. Another approach for obtaining solutions is the vertex operator for the ultradiscrete KdV equation, which is proposed by the author in \cite{N}. This approach is believed to be closely related to certain types of symmetries for this system.

In this paper, we propose a vertex operator for the non-autonomous ultradiscrete KP equation and various ultradiscrete soliton equations obtained by reduction. In section 2 we first propose a recursive representation of the soliton solutions of the non-autonomous ultradiscrete KP equation. In section 3, we propose the vertex operator as an operator representation of the recursive one. In section 4, we present various reductions of this equation and discuss their vertex operators and solutions. Finally, in section 5, we give some concluding remarks.

\section{Recursive expression for the solution of the ultradiscrete KP equation}

The non-autonomous ultradiscrete KP equation is written as
\begin{equation}\label{bilinear}
\fl\quad	T_{l, m+1, n} + T_{l+1, m, n+1} = \max \big(\ T_{l+1, m, n} + T_{l, m+1, n+1} - 2R_{n}, T_{l, m, n+1} + T_{l+1, m+1, n}\ \big),
\end{equation}
where $R_{n} \ge 0$ depends only on $n$.

\begin{Thm}\label{Thm1}
The function $T^{(N)}_{l, m, n}$ expressed as
\begin{equation}\label{exformula}
	T^{(N)}_{l, m, n} =	\cases{\max \big( T_{l, m, n}^{(N-1)}, 2 \eta_{N} + T^{(N-1)}_{l-1, m+1, n} \big) &($N \ge 1$)\\0&($N = 0$)\\}
\end{equation}
solves equation (\ref{bilinear}) for $\eta_{N}$ given by
\begin{equation}\label{defeta}
	\eta_{N} = C_{N} + l P_{N} - m Q_{N} -  \sum_{0}^{n} \Omega_{N, d}.
\end{equation}
Here, $\sum_{i}^{j} \Omega_{N,d}$ stands for
\begin{equation}
 \sum_{i}^{j} \Omega_{N, d} = \cases{\sum_{d=i+1}^{j} \Omega_{N, d}&($i<j$)\\0&($i=j$)\\-\sum_{d=j+1}^{i} \Omega_{N, d} &($i>j$)\\},
\end{equation}
and the parameters $P_{i}, Q_{i}$ and $ \Omega_{i, n} (i=1,\ldots, N)$ satisfy the relations:
\begin{eqnarray}
	P_{N} &\ge P_{N-1} \ge \ldots \ge P_{1} \ge 0 \label{condp} \\
	Q_{N} &\ge Q_{N-1} \ge \ldots \ge Q_{1} \ge 0 \label{condq} \\
	\Omega_{i, n} &= \min ( Q_{i}, R_{n-1} ). \label{defomega}
\end{eqnarray}
\end{Thm}

\begin{Lem}\label{Lem1}
Let
\begin{equation}
	H^{(N)}_{l, m, n} = T^{(N)}_{l, m+j+1, n+k} + T^{(N)}_{l+i+1, m, n} - T^{(N)}_{l+1, m+j, n+k} - T^{(N)}_{l+i, m+1, n}
\end{equation}
for $i, j, k$ such that
\begin{equation}\label{condvanish}
	i P_{N} + j Q_{N} + \sum_{n}^{n+k} \Omega_{N, d} \ge \ldots \ge i P_{1} + j Q_{1} + \sum_{n}^{n+k} \Omega_{1, d} \ge 0.
\end{equation}
Then it holds that
\begin{equation}
	H^{(N)}_{l,m,n} \le 2(i P_{N} + j Q_{N} + \sum_{n}^{n+k} \Omega_{N, d} ).
\end{equation}
\end{Lem}

\begin{Proof}
By employing the inequality
\begin{equation}\label{maxtrineq}
	\max ( a, b ) - \max ( c, d ) \le \max ( a - c, b - d ),
\end{equation}
we obtain
\begin{eqnarray}
\fl\quad T^{(N)}_{l, m+j+1, n+k} - T^{(N)}_{l+i, m+1, n} \le \max \big( T^{(N-1)}_{l, m+j+1, n+k} - T^{(N-1)}_{l+i, m+1, n}, \nonumber\\ 
		\qquad\quad -2(i P_N - j Q_N - \sum_{n}^{n+k} \Omega_{N, d}) + T^{(N-1)}_{l-1, m+j+2, n+k} - T^{(N-1)}_{l+i-1, m+2, n} \big)
\end{eqnarray}
\begin{eqnarray}
\fl\quad T^{(N)}_{l+i+1, m, n} - T^{(N)}_{l+1, m+j, n+k} \le \max \big( T^{(N-1)}_{l+i+1, m, n} - T^{(N-1)}_{l+1, m+j, n+k}, \nonumber\\ 
	\qquad\quad	2(i P_N + j Q_N + \sum_{n}^{n+k} \Omega_{N, d}) + T^{(N-1)}_{l+i, m+1, n} - T^{(N-1)}_{l, m+j+1, n+k} \big)
\end{eqnarray}
Adding the inequalities yields
\begin{eqnarray}
\fl\quad H^{(N)}_{l, m, n} \le \max \big( H^{(N-1)}_{l, m, n}, \ 2(i P_N + j Q_N + \sum_{n}^{n+k} \Omega_{N, d} ), \nonumber\\
								-2(i P_N + j Q_N + \sum_{n}^{n+k} \Omega_{N, d} ) + H^{(N-1)}_{l, m, n} + H^{(N-1)}_{l-1, m+1, n}, 	\ H^{(N-1)}_{l-1, m+1, n} \big)
\end{eqnarray}
Taking into account the relations $i P_{N} + j Q_{N} + \sum_{n}^{n+k} \Omega_{N, d} \ge i P_{N-1} + j Q_{N-1} + \sum_{n}^{n+k} \Omega_{N-1, d}$, it can be shown inductively that the four arguments in this maximum are all less than $2(i P_N + j Q_N + \sum_{n}^{n+k} \Omega_{N, d} )$. \hfill \opensquare
\end{Proof}

\begin{Lem}\label{Lem2}
Let
\begin{equation}
	H'^{(N)}_{l, m, n} = T^{(N)}_{l, m, n+1} + T^{(N)}_{l, m+2, n} - T^{(N)}_{l, m+1, n} - T^{(N)}_{l, m+1, n+1}.
\end{equation}
One then has
\begin{equation}\label{lem2ineq}
	H'^{(N)}_{l, m, n} \le 2(Q_N - \Omega_{N, n+1})
\end{equation}
when one requires that the $T^{(i)}_{l, m, n} (i=1, \ldots, N)$ are solutions of (\ref{bilinear}). Especially for $\Omega_{N,n} = Q_{N}$, the inequality (\ref{lem2ineq}) becomes an equality, i.e.: $H'^{(N)}_{l,m,n} = 0$ .
\end{Lem}

\begin{Proof}
When $\Omega_{N,n} = R_{n+1}$, we obtain by virtue of the inequality (\ref{maxtrineq}):
\begin{eqnarray}
	H'^{(N)}_{l, m, n} \le \max \big( H'^{(N-1)}_{l, m, n}, H'^{(N-1)}_{l-1, m+1, n}, \nonumber\\
	2( Q_N - \Omega_{N, n+1} ) + T^{(N-1)}_{l-1, m+1, n+1} + T^{(N-1)}_{l, m+2, n} - T^{(N-1)}_{l-1, m+2, n} - T^{(N-1)}_{l, m+1, n+1}, \nonumber\\
	 -2( Q_N - \Omega_{N, n+1} ) + H^{(N-1)}_{l-1, m, n+1}\big|_{(i,j,k)=(0,1,-1)} + H'^{(N-1)}_{l-1,m+1,n}\big).
\end{eqnarray}
However, $T^{(N-1)}_{l-1, m+1, n+1} + T^{(N-1)}_{l, m+2, n} - T^{(N-1)}_{l-1, m+2, n} - T^{(N-1)}_{l, m+1, n+1} \le 0$ because $T^{(N-1)}_{l,m,n}$ satisfies (\ref{bilinear}). It can then be shown inductively that all arguments in the maximum are less than $2(Q_N - \Omega_{N, n+1})$. 

On the other hand, when $\Omega_{N,n} = Q_{N}$, by virtue of (\ref{condq}), $T^{(N-1)}_{l, m+1, n}$ is equal to $T^{(N-1)}_{l, m, n+1}$ for all $l, m$ because
\begin{equation}
\fl \qquad	C_{i} + l P_{i} - (m+1) Q_{N} -  \sum_{0}^{n} \Omega_{N, d} = C_{N} + l P_{N} - m Q_{N} -  \sum_{0}^{n+1} \Omega_{N, d}
\end{equation}
for all $i=1, \ldots, N$. We thus obtain that
\begin{equation}
	H'^{(N)}_{l, m, n} = (T^{(N)}_{l, m, n+1} - T^{(N)}_{l, m+1, n}) + (T^{(N)}_{l, m+2, n} - T^{(N)}_{l, m+1, n+1}) = 0.
\end{equation}\hfill \opensquare
\end{Proof}

\begin{Lem}\label{Lem3}
Let
\begin{equation}
	H''^{(N)}_{l, m, n} = T^{(N)}_{l, m, n+1} + T^{(N)}_{l+2, m, n} - T^{(N)}_{l+1, m, n} - T^{(N)}_{l+1, m, n+1}.
\end{equation}
One then has
\begin{equation}
	H''^{(N)}_{l, m, n} \le 2 P_N
\end{equation}
when all of $T^{(i)}_{l, m, n} (i=1, \ldots, N)$ are solutions of (\ref{bilinear}).
\end{Lem}

\begin{Proof}
The proof is essentially the same as that of Lemma \ref{Lem2} when $\Omega_{N, n} = R_{n+1}$. \hfill \opensquare
\end{Proof}

\bigskip

We now have all the necessary lemmas at our disposal and proceed to the proof of theorem \ref{Thm1}.

\bigskip

We shall prove the theorem inductively. It is clear that $T^{(0)}_{l, m, n}$ solves equation (\ref{bilinear}) because of the non-negativity of $R_n$. Now, let us assume that the theorem holds at $1, \ldots, N-1$. By substituting (\ref{exformula}) in equation (\ref{bilinear}), each contribution can be written as
\begin{eqnarray} \label{ff1}
\fl \quad	T^{(N)}_{l, m+1, n} + T^{(N)}_{l+1, m, n+1} = &\max \big(\ T^{(N-1)}_{l, m+1, n} + T^{(N-1)}_{l+1, m, n+1}, \nonumber\\
	&\quad 2(P_{N} - \Omega_{N, n+1}) + 2 \eta_{N} + T^{(N-1)}_{l, m+1, n} + T^{(N-1)}_{l, m+1, n+1}, \nonumber\\
	&\quad -2 Q_{N} + 2 \eta_{N} + T^{(N-1)}_{l-1, m+2, n} + T^{(N-1)}_{l+1, m, n+1}, \nonumber\\
	&\quad 4 \eta_{N} + 2(P_{N} - Q_{N} - \Omega_{N, n+1}) + T^{(N-1)}_{l-1, m+2, n} + T^{(N-1)}_{l, m+1, n+1} \big),
\end{eqnarray}
for the left hand side of (\ref{bilinear}), and
\begin{eqnarray} \label{ff2}
\fl \quad	T^{(N)}_{l+1, m, n} + T^{(N)}_{l, m+1, n+1} = &\max \big( T^{(N-1)}_{l, m, n} + T^{(N-1)}_{l, m+1, n+1}, \nonumber \\
	&\quad 2 P_{N} + T^{(N-1)}_{l, m+1, n} + T^{(N-1)}_{l-1, m+1, n+1}, \nonumber\\
	&\quad -2 ( Q_{N} + \Omega_{N, n+1} ) + T^{(N-1)}_{l+1, m, n} + T^{(N-1)}_{l, m+2, n+1}, \nonumber\\
	&\quad 4 \eta_{N} + 2(P_{N} - Q_{N} - \Omega_{N, n+1}) + T^{(N-1)}_{l, m+1, n} + T^{(N-1)}_{l-1, m+2, n+1} \big) \\
\label{ff3}
\fl \quad T^{(N)}_{l, m, n+1} + T^{(N)}_{l+1, m+1, n} = &\max \big( T^{(N-1)}_{l, m, n+1} + T^{(N-1)}_{l+1, m, n}, \nonumber\\
	&\quad 2 (P_{N} - Q_{N}) + T^{(N-1)}_{l, m, n+1} + T^{(N-1)}_{l, m+2, n}, \nonumber\\
	&\quad -2 \Omega_{N, n+1} + T^{(N-1)}_{l-1, m+1, n+1} + T^{(N-1)}_{l+1, m+1, n}, \nonumber\\
	&\quad 4 \eta_{N} + 2(P_{N} - Q_{N} - \Omega_{N, n+1}) + T^{(N-1)}_{l-1, m+1, n+1} + T^{(N-1)}_{l, m+2, n} \big)
\end{eqnarray}
for the right hand side. In these expressions it looks as if each of the maximum operations in (\ref{ff1})--(\ref{ff3}) has four arguments. However, by virtue of Lemma \ref{Lem1}, the third argument in (\ref{ff1}) and (\ref{ff2}) cannot yield the maximum because it is always less than the second argument.

Then, the relevant arguments of the maximum in (\ref{ff1}) are in fact
\begin{eqnarray}
	&T^{(N-1)}_{l, m+1, n} + T^{(N-1)}_{l+1, m, n+1} \label{lhs1} \\
	&2 \eta_{N} + 2(P_{N} - \Omega_{N, n+1}) + T^{(N-1)}_{l, m+1, n} + T^{(N-1)}_{l, m+1, n+1} \label{lhs2} \\
	&4 \eta_N + 2(P_{N} - Q_{N} - \Omega_{N, n+1}) + T^{(N-1)}_{l-1, m+2, n} + T^{(N-1)}_{l, m+1, n+1} \label{lhs3}
\end{eqnarray}
and those in the maximum of the contributions in (\ref{ff2}), (\ref{ff3}), as they appear in the right hand side of equation (\ref{bilinear}):
\begin{eqnarray}
\fl \max ( T^{(N-1)}_{l+1, m, n} + T^{(N-1)}_{l, m+1, n+1} - 2R, T^{(N-1)}_{l, m, n+1} + T^{(N-1)}_{l+1, m, n} ) \label{rhs1} \\
\fl	2 \eta_{N} + \max ( 2 P_{N} - 2 R_{n} + T^{(N-1)}_{l, m+1, n} + T^{(N-1)}_{l, m+1, n+1}, \nonumber\\
\fl \quad 2(P_N - Q_N) + T^{(N-1)}_{l, m, n+1} + T^{(N-1)}_{l, m+2, n}, -2 \Omega_{N, n+1} + T^{(N-1)}_{l-1, m+1, n+1} + T^{(N-1)}_{l+1, m+1, n} ) \label{rhs2} \\
\fl	4 \eta_N + 2(P_N - Q_N - \Omega_{N, n+1}) \nonumber\\
\fl \quad + \max ( T^{(N-1)}_{l, m+1, n} + T^{(N-1)}_{l-1, m+2, n+1} - 2R, T^{(N-1)}_{l-1, m+1, n+1} + T^{(N-1)}_{l, m+2, n} ). \label{rhs4}
\end{eqnarray}
Here, (\ref{lhs1}) and (\ref{lhs3}) are identical to (\ref{rhs1}) and (\ref{rhs4}) because  by assumption, $T^{(N-1)}_{l, m, n}$ solves the equation (\ref{bilinear}). 

By subtracting (\ref{lhs2}) from (\ref{rhs2}), we obtain
\begin{equation}\label{mid1}
\fl \quad \quad \quad \max \big( 2(\Omega_{N, n+1} - R_{n}), 2(\Omega_{N, n+1} - Q_{N}) + H'^{(N-1)}_{l,m,n}, -2 P_{N} + H''^{(N-1)}_{l,m,n} \big).
\end{equation}
The third argument of this maximum is non-positive by virtue of Lemma \ref{Lem3}.

In the case $\Omega_{N,n+1} = Q_{N} \le R_{n}$, $\Omega_{N-1}$ has to be equal to $Q_{N-1}$ due to condition (\ref{condq}). Then, the first argument in (\ref{mid1}) is also non-positive and the second argument is $0$, due to Lemma \ref{Lem2}.

In the case $\Omega_{N,n+1}= R_{n}$, the first argument in the maximum in (\ref{mid1}) is $0$ and the second argument is non-positive by virtue of Lemma \ref{Lem2}. Thus, (\ref{mid1}) is equal to $0$, in all possible cases (i.e., $\Omega_{N,n+1} = Q_{N}$ or $\Omega_{N,n+1} = R_{n}$).

We have therefore shown that all arguments of the maximum in (\ref{ff1}) which constitutes the left hand side of (\ref{bilinear}), have an equivalent counterpart among (\ref{rhs1}), (\ref{rhs2}), (\ref{rhs4}), i.e. among the three arguments that contribute to the right hand side of (\ref{bilinear}). Hence, (\ref{bilinear}) is satisfied.\hfill \opensquare

Please note that the proof allows for the possibility that, at different values of $n$, $\Omega_{N,n+1}$ satisfies different equalities ($\Omega_{N,n+1}=R_{n}$ or $\Omega_{N,n+1}=Q_{N}$ for different $n$), because the shift of the independent variables induced by (\ref{exformula}) affects only $l$ and $m$, not $n$. 

\section{Vertex operator for the KP equation}

In this section we propose an alternative representation of the $N$-soliton solutions, generated by a vertex operator $X$.

The $0$-soliton solution $T(;;)$ is written as:
\begin{equation}
	T(;;) := 0
\end{equation}
whereas the $N+1$-soliton solution is generated from the $N$-soliton solution \newline $T(P_1, \ldots, P_{N};Q_{1}, \ldots, Q_{N}; C_{1}, \ldots, C_{N})$ (written as $T(\boldsymbol{P}; \boldsymbol{Q}; \boldsymbol{C})$ for brevity) by
\begin{eqnarray}\label{symexformula}
\fl\qquad	X(P_{N+1}, Q_{N+1}, C_{N+1}) T(\boldsymbol{P}; \boldsymbol{Q}; \boldsymbol{C}) \nonumber\\
	:= \max ( T(\boldsymbol{P}; \boldsymbol{Q}; \boldsymbol{C}), 2\eta_{N+1} + T(\boldsymbol{P}; \boldsymbol{Q}; \boldsymbol{C}-\boldsymbol{A}_{N+1}) ) \\
	=: T(P_1, \ldots, P_{N}, P_{N+1};Q_{1}, \ldots, Q_{N}, Q_{N+1}; C_{1}, \ldots, C_{N}, C_{N+1}), 
\end{eqnarray}
where the parameters $P_{N+1}, Q_{N+1}$ in the vertex operator $X$ must satisfy 
\begin{equation}
	(P_{i}-P_{N+1})(Q_{i}-Q_{N+1}) \ge 0.
\end{equation}
The phase factor $\eta_{N+1}$ is the same as in (\ref{defeta}), and the interaction terms \newline $\boldsymbol{A}_{N+1} = {}^t (A_{N+1,1}, \ldots, A_{N+1, N})$ are
\begin{equation}
	A_{i, j} = \min (P_i, P_j) + \min (Q_i, Q_j).
\end{equation}

\begin{Prop} \label{sym}
The action of the operator $X$ is commutative.
\end{Prop}

\begin{Proof}
By calculating $X(\Omega_b, \eta_b) X(\Omega_a, \eta_a) F({\boldsymbol{\Omega}}; {\boldsymbol{\eta}})$ directly, we obtain
\begin{eqnarray}
\fl\qquad	X(P_b, Q_b, C_b) X(P_a, Q_a, C_a) T(\boldsymbol{P}; \boldsymbol{Q}; \boldsymbol{C}) \nonumber\\
	\!\!\!\!\!\!\!\! = \max \big( T(\boldsymbol{P}; \boldsymbol{Q}; \boldsymbol{C}), 2 \eta_{b} + T(\boldsymbol{P}; \boldsymbol{Q}; \boldsymbol{C} - \boldsymbol{A}_{b}), \nonumber\\
	  2 \eta_{a} + T(\boldsymbol{P}; \boldsymbol{Q}; \boldsymbol{C} - \boldsymbol{A}_{a}), 2 \eta_{a} + 2 \eta_{b} - 2 A_{b, a} + T(\boldsymbol{P}; \boldsymbol{Q}; \boldsymbol{C} - \boldsymbol{A}_{a} - \boldsymbol{A}_{b}) \big).
\end{eqnarray}
From this relation it is clear that interchanging the subscripts $a$ and $b$ does not change the overall value of the maximum. \hfill $\square$
\end{Proof}

Rewriting this proposition yields the following corollary:
\begin{Cor}\label{cor1}
The $N$-soliton solution $T(\boldsymbol{P}; \boldsymbol{Q}; \boldsymbol{C})$ is invariant under the permutation of its parameters, i.e.:
\begin{eqnarray}
\!\!\!\!\!\! T(P_1, \ldots, P_{N};Q_{1}, \ldots, Q_{N}; C_{1}, \ldots, C_{N}) \nonumber\\
	= T(P_{\sigma(1)}, \ldots, P_{\sigma(N)};Q_{\sigma(1)}, \ldots, Q_{\sigma(N)}; C_{\sigma(1)}, \ldots, C_{\sigma(N)}) \quad ( \sigma \in S_{N})
\end{eqnarray}
\end{Cor}

By virtue of corollary \ref{cor1}, we can fix the labels of the parameters as in (\ref{condp}), (\ref{condq}) without loss of generality. By virtue of this ordering, the phase shifts in $A_{i, j}$ in the definition (\ref{symexformula}) simplify to
\begin{equation}
	\min(P_{i}, P_{N}) = P_{i},\ \min(Q_{i}, Q_{N}) = Q_{i} \quad (i=1,\ldots,N-1).
\end{equation}
It should be noted that the phase shifts $\boldsymbol{\eta} \to \boldsymbol{\eta} + \boldsymbol{P}$ and $\boldsymbol{\eta} \to \boldsymbol{\eta} + \boldsymbol{Q}$ are equivalent to shifts on the independent variables $l \to l + 1$ and $m \to m - 1$, which shows that   $T(\boldsymbol{P}; \boldsymbol{Q}; \boldsymbol{C})$ is equivalent to $T^{(N)}_{l, m, n}$.

\section{Reduction to various ultradiscrete soliton equations}

In this section we present some examples of reductions of the ultradiscrete KP equation to $1+1$ dimensional ultradiscrete equations and we give the vertex operators for these equations.

\subsection{The Box and Ball System and its varieties}

By restricting $T_{l,m,n}$ to
\begin{equation}\label{redhKdV}
	T_{l, m, n} = F^{l-Mm}_{n}
\end{equation}
and denoting $s=l-Mm$ and $n=j$, the non-autonomous ultradiscrete KP equation (\ref{bilinear}) is reduced to the so-called non-autonomous ultradiscrete hungry KdV equation:
\begin{equation}\label{uhKdVbilinear}
	F^{s+M+1}_{j+1} + F^{s}_{j} = \max ( F^{s+M+1}_{j} + F^{s}_{j+1} - 2R_{j}, F^{s+1}_{j} + F^{s+M}_{j+1} ).
\end{equation}
By means of the dependent variable transformation
\begin{equation}
	B^{t}_{i,j} = \frac{1}{2} ( F^{s+1}_{j} + F^{s}_{j+1} - F^{s+1}_{j+1} - F^{s}_{j} ),
\end{equation}
and denoting $s=Mt+i$, (\ref{uhKdVbilinear}) is transformed into
\begin{equation}
	B^{t+1}_{i,j} = \min \Big( R_{j} - \sum_{k=1}^{i-1} B^{t+1}_{k,j} - \sum_{k=i}^{M} B^{t}_{k,j}, \sum_{n=-\infty}^{j-1} ( B^{t}_{i,n} - B^{t+1}_{i,n} ) \Big),
\end{equation}
which describes the dynamics of a Box and Ball System with $M$ kinds of balls, as presented in \cite{TTM}. This system is required to satisfy the following boundary conditions:
\begin{equation}\label{bchKdV}
	B^{t}_{i,j} = 0	\quad	\mbox{for}	\quad j\ll 0
\end{equation}
In particular, in the case of $M=1$ it reduces to an extension of the standard BBS \cite{TS}, with variable size of boxes at each site.

In our representation (\ref{symexformula}), the reduction (\ref{redhKdV}) is equivalent to the parameter restriction:
\begin{equation}
	M P_{N} = Q_{N}.
\end{equation}
It should be noted that our representation satisfies the boundary condition (\ref{bchKdV}) because the first argument of $\max$ in (\ref{symexformula}) is never chosen for sufficiently small $j$.

Then, the vertex operator for (\ref{uhKdVbilinear}) can be written as
\begin{eqnarray}\label{symexformulahKdV}
\fl\qquad	X(P_{N+1}, C_{N+1}) T(\boldsymbol{P}; \boldsymbol{C}) \nonumber\\
	:= \max ( T(\boldsymbol{P}; \boldsymbol{C}), 2\eta_{N+1} + T(\boldsymbol{P}; \boldsymbol{C}-\boldsymbol{A}_{N+1}) ),
\end{eqnarray}
where the phase factor $\eta_{N+1}$ is 
\begin{equation}
	\eta_{N} = C_{N} + s P_{N} -  \sum_{0}^{j} \Omega_{N, d},
\end{equation}
and $\Omega_{N,j}$ and the interaction terms $A_{i, j}$ are expressed as
\begin{equation}
	\Omega_{N,j} = \min ( R_{j-1}, M P_{N}), \quad  A_{i, j} = (M+1) \min (P_i, P_j).
\end{equation}

\subsection{The ultradiscrete Toda equation}

By restricting $T_{l,m,n}$ to
\begin{equation}\label{redToda}
	T_{l, m, n} = F^{l+n}_{m+n},
\end{equation}
$R_n = const.$ and denoting $t=l+n$ and $s=m+n$, (\ref{bilinear}) is reduced to the ultradiscrete Toda equation:
\begin{equation}\label{uTodabilinear}
	F^{t}_{s+1} + F^{t+2}_{s+1} = \max ( F^{t+1}_{s+2} + F^{t+1}_{s} - 2R, 2 F^{t+1}_{s+1} ) 
\end{equation}
By means of the dependent variable transformation
\begin{equation}
	U^{t}_{s} = \frac{1}{2} ( F^{t}_{s+2} - 2 F^{t}_{s+1} + F^{t}_{s} ),
\end{equation}
(\ref{uTodabilinear}) is transformed into
\begin{equation}
\fl\	U^{t+2}_{s+1} - 2 U^{t+1}_{s+1} + U^{t}_{s+1} = \max ( U^{t+1}_{s+2} - R, 0 ) - 2 \max ( U^{t+1}_{s+1} -R, 0 ) + \max ( U^{t+1}_{s} - R, 0 ),
\end{equation}
which describes the dynamics of the Toda type cellular automaton presented in \cite{MSTTT}.

In our representation (\ref{symexformula}), the reduction (\ref{redToda}) is equivalent to the parameter restriction:
\begin{equation}
	\Omega_{N} = Q_{N} - P_{N} \qquad  \mbox{i.e.} \qquad P_{N} = Q_{N} - \Omega_{N} = \max ( Q_{N} - R, 0 )
\end{equation}

The vertex operator of (\ref{uTodabilinear}) can be expressed as
\begin{eqnarray}\label{symexformulaToda}
\fl\qquad	X(P_{N+1}, C_{N+1}) T(\boldsymbol{P}; \boldsymbol{C}) \nonumber\\
	:= \max ( T(\boldsymbol{P}; \boldsymbol{C}), 2\eta_{N+1} + T(\boldsymbol{P}; \boldsymbol{C}-\boldsymbol{A}_{N+1}) ),
\end{eqnarray}
where the phase factor $\eta_{N+1}$ is 
\begin{equation}
	\eta_{N} = C_{N} + t \max ( Q_{N} - R, 0 )- s Q_{N},
\end{equation}
and the interaction term $A_{i, j}$ is written as
\begin{equation}
	A_{i, j} = \min ( Q_{i}, Q_{j} ) + \max ( \min ( Q_{i}, Q_{j} ) - R, 0 ) 
\end{equation}

\section{Concluding Remarks}
In this paper, we proposed a recursive representation of the $N$-soliton solutions and vertex operators for the ultradiscrete KP equation. We also proposed expressions for various ultradiscrete equations, obtained by reduction from the KP equation.

In fact, the vertex operator approach is closely related to the existence of certain symmetry algebras for integrable systems and the exact relation of our ultradiscrete operator to the symmetries of ultradiscrete systems is an especially interesting problem we want to address in the future.

Because it uses simple shift and $\max$ operators and not the usual algebraic or combinatorial methods, our representation also has the potential to describe solutions different from the solitonic ones. It is an interesting problem to describe the full class of solutions these equations admit.


\section*{References}
\begin{thebibliography}{10}

\bibitem{NTS}
A.~Nagai, T.~Tokihiro, and J.~Satsuma.
\newblock Conserved quantities of box and ball system.
\newblock {\em Glasg. Math. J.}, 43A:91--97, 2001.

\bibitem{TS}
D.~Takahashi and J.~Satsuma.
\newblock A soliton cellular automaton.
\newblock {\em J. Phys. Soc. Jpn.}, 59:3514--3519, 1990.

\bibitem{TTMS}
T.~Tokihiro, D.~Takahashi, J.~Matsukidaira, and J.~Satsuma.
\newblock {From Soliton Equations to Integrable Cellular Automata through a
  Limiting Procedure}.
\newblock {\em Phys. Rev. Lett.}, 76:3247--3250, 1996.

\bibitem{WTS}
R.~Willox, T.~Tokihiro, and J.~Satsuma.
\newblock Nonautonomous discrete integrable systems.
\newblock {\em Chaos, Solitons and Fractals}, 11:121--135, 2000.

\bibitem{TTM}
T.~Tokihiro, D.~Takahashi, and J.~Matsukidaira.
\newblock {Box and ball system as a realization of ultradiscrete nonautonomous
  KP equation}.
\newblock {\em J. Phys. A: Math. Gen.}, 33:607--619, 2000.

\bibitem{SH}
N.~Shinzawa and R.~Hirota.
\newblock {The B\"acklund transformation equations for the ultradiscrete KP
  equation}.
\newblock {\em J. Phys. A: Math. Gen.}, 36:4667--4675, 2003.

\bibitem{TH}
D.~Takahashi and R.~Hirota.
\newblock Ultradiscrete soliton solution of permanent type.
\newblock {\em J. Phys. Soc. Jpn.}, 76:104007, 2007.

\bibitem{Nh}
H.~Nagai.
\newblock {A new expression of soliton solution to the ultradiscrete Toda
  equation}.
\newblock {\em J. Phys. A: Math. Theor.}, 41:235204 (12pp), 2008.

\bibitem{N}
Y.~Nakata.
\newblock {Vertex operator for the ultradiscrete KdV equation}.
\newblock {\em J. Phys. A: Math. Theor.}, 42:412001 (6pp), 2009.

\bibitem{MSTTT}
J.~Matsukidaira, J.~Satsuma, D.~Takahashi, T.~Tokihiro, and M.~Torii.
\newblock {Toda-type cellular automaton and its $N$-soliton solution}.
\newblock {\em Phys. Lett. A}, 225:287--295, 1997.

\end{thebibliography}
\end{document}